\documentclass[preprint]{revtex4}
\usepackage{graphicx,epsfig,amssymb}

\begin{document}

\title{Human housekeeping genes are compact}
 
\author{Eli Eisenberg and Erez Y. Levanon}
\affiliation{Compugen Ltd., 72 Pinchas Rosen Street, Tel Aviv 69512, Israel}

\begin{abstract}
We identify a set of 575 human genes that are
expressed in all conditions tested in a publicly available
database of microarray results. Based on this common
occurrence, the set is expected to be rich in ``housekeeping''
genes, showing constitutive expression in all tissues.
We compare selected aspects of their genomic structure
with a set of background genes. We find that the
introns, untranslated regions and coding sequences
of the housekeeping genes are shorter, indicating a
selection for compactness in these genes.
\end{abstract}

\maketitle 

The amazing diversity of the human body stems from the
different expression patterns of genes in different tissues.
Although most genes show constitutive expression in only
a subset of tissues, some gene products are required for the
maintenance of the basal cellular function and are
constitutively found in all human cells. These genes are
called housekeeping genes (HK genes) \cite{1}. HK genes can
be used to calibrate measurements of gene expression \cite{2}.
They might also help to define the minimal gene complement
needed for a human cell \cite{1}. Several attempts have been
made recently to define the complete set of HK genes \cite{3,4}.

Microarrays are often used to identify sets of genes that
are expressed either ubiquitously or in specific tissues or
conditions. However, the technique is technically demanding
and prone to artifacts, so independent evidence is often
required to confirm the results. In principle, identifying
the set of HK genes using microarray data is straightforward;
one need only look for genes that are expressed in all
tissues and all experimental conditions. Employing such
an approach has so far resulted in two lists of HK genes
\cite{3,4}. However, problems in probe design, measurement
noise and other artifacts introduce inevitable errors in
such lists. Because a northern blot experiment for each
gene in each tissue is impractical, an independent test is
needed to validate any list of HK genes. Here, we report a
validation test that uses a recently discovered property of
highly expressed genes.

The transcription process is both slow and costly; it
takes 50 milliseconds \cite{5,6} and two ATP molecules \cite{7}
approximately to transcribe a nucleotide. This might be
expected to provide selective pressure to make genes as
short as functionally possible. The more copies of a gene
required for the organism, the stronger this pressure
should be. The first demonstration of this principle \cite{8}
showed that genes with a large number of expressed
sequence tags (ESTs) in public libraries (and hence most
mRNAs) have a significantly shorter average intron length
than those with fewer ESTs.

Here, an implication of this principle is used to validate
a set of HK genes. The HK genes, which are transcribed in
all somatic cells and under all circumstances, are by
nature highly expressed, and therefore should be selected
to have shorter introns. We used a recently published
database of microarray experiments \cite{9} to identify a set of
HK genes. As a further validation step, we checked the
Gene Ontology (GO) annotation of these genes. We
compared the structure of the HK genes with all other
genes, and not only the introns, but all parts of the HK
genes were found to be, on average, shorter than other
genes. In particular, the untranslated regions and the
translated proteins are all shorter in the HK genes.

\section{Assignment of housekeeping genes}

\begin{figure}
\includegraphics[width=2.5in]{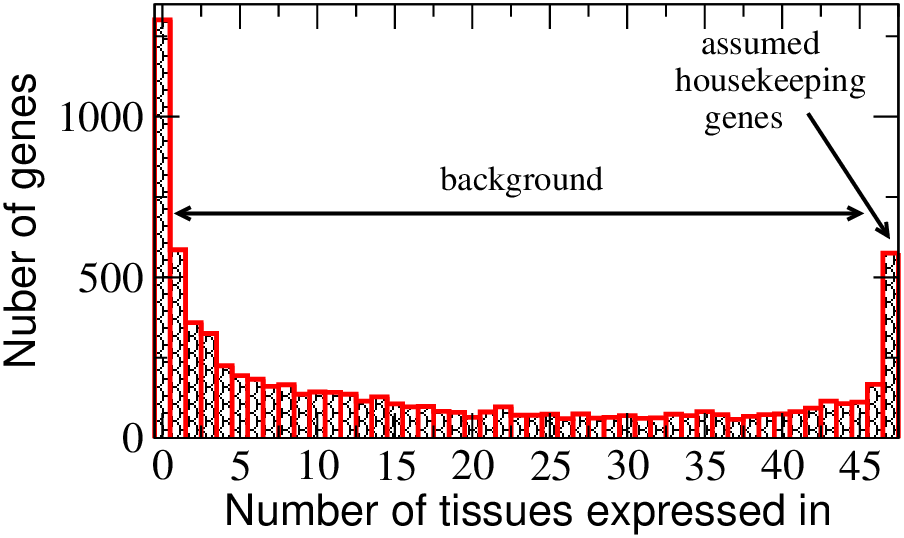}
\caption{The distribution of 7500 RefSeq genes represented on 
the microarray as a function of the number of tissues they express in. 
Each bin gives the number of genes
expressed in M out of 47 different tissues. The M=47 bin corresponds to 
the housekeeping genes, expressed in all tissues.}
\end{figure}

A recently published database provides microarray
expression data for Affymetrix U95A chip, containing
12,600 probes, and hybridized to 101 different samples \cite{9}
from 47 different human tissues and cell lines. These
samples are mainly from the normal human physiological
state, and therefore this dataset provides a description of
the normal mammalian transcriptome.

We calculated the distribution of the number of different
tissues in which a gene is expressed. Discarding probes for
which the associated gene was not represented in the
RefSeq database \cite{10}, and unifying all probes measuring
the same gene (ignoring the potential differences among
splice variants) yielded probes representing 7500 human
genes. The experiments measuring replicates of the same
biological condition were averaged to reduce the measurement
noise, resulting in 47 data points per probe. We
considered that a probe was expressed in a certain
condition if its average reading was above a certain cutoff
value. The results were not sensitive to the exact cut-off
value, and we chose 200 standard Affymetrix averagedifference
units, considered to be a conservative cut-off
value for determining gene presence \cite{9}. This is also the
trimmed average expression level in each tissue in
accordance with the standard Affymetrix normalization
procedure \cite{11,12}. Thus, our HK genes are expressed in all
tissues at an above-average level.

A histogram (Fig. 1) of the number of genes expressed
in exactly M of the 47 tissues shows a clear tendency for
frequency to decrease as M increases. However, a
substantial number of genes (575), belong to the class of
genes that are expressed in all tissues. Because their
number is far greater than expected based on the general
trend described above, we assumed this class to be rich in
HK genes, and considered it to be the set of HK genes.

It is noteworthy that the genes in our HK list tend to
have an average expression significantly higher than other
genes; the geometric mean expression of our HK genes is
1200 in Affymetrix average difference units, whereas that
of other genes is 150. The difference cannot be accounted
for by the cutoff used to define the HK genes, and is not a
result of a bias due to inclusion of genes expressed in a few
tissues only (data not shown).

Two additional tests were conducted to validate this set.
First, a study of the GO annotation \cite{13} of these genes
revealed the set is rich in metabolic proteins (24\%) and
RNA-interacting proteins (19\%, mostly ribosomal proteins).
Second, we compiled a list of 18 well-established
HK genes commonly used for quantitative PCR calibration
\cite{14,15}, and checked our list against it. We found 13 of
the 18 genes in our list, and the other five were not
represented on the microarray (see Table in Supplementary
Information at http://www.compugen.co.il/supp\_info/Housekeeping\_genes.html).

\section{Length analysis of HK genes}

\begin{figure}[t]
\includegraphics[width=2.5in]{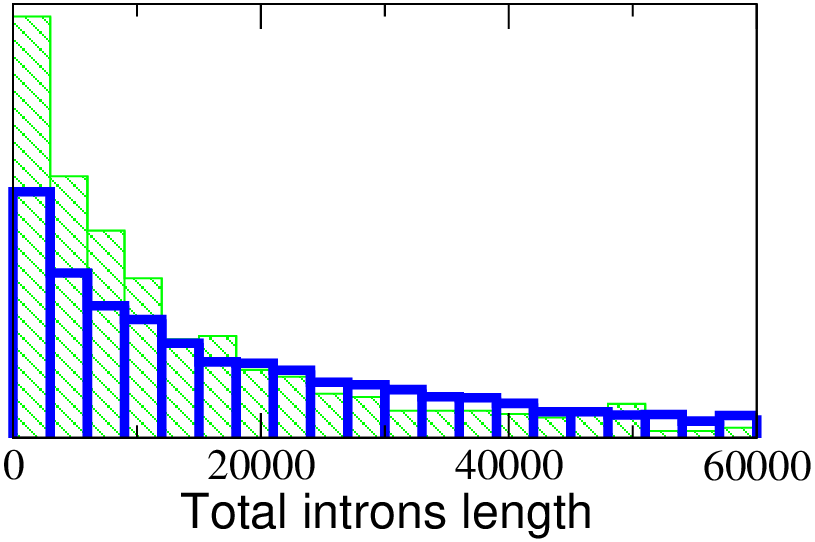}
\caption{A Histogram of the total length of introns. Green bars, HK genes; 
blue bars, non-HK genes.}
\end{figure}

\begin{table}[b]

\caption{{\bf Human housekeeping genes are compact.} 
Comparison of structure of housekeeping (HK) 
genes versus non-HK genes. For each case the first 
line gives the average value, s.e.m, and the second line gives the median.
For the average intron and exon lengths, 
all introns and exons belonging to the relevant set were 
included; the number appears in parentheses. The P-value was calculated
using the Mann-Whitney test. UTR, untranslated region.
}

\begin{tabular}{llll}
& & &\\
& {\bf HK genes (n=532)}&{\bf non-HK (n=5404)}& {\bf P-value}\\
Average intron length&$2573\pm 145$(n=4353)\ \ \ \ \ \ &$5025\pm 71$(n=57447)\ \ \ \ \ \ &$4\times 10^{-130}$\\
&672&1365& \\
Total intron length&$21050\pm 1781$&$53418\pm 1425$&$7\times 10^{-28}$\\
&9293&20804& \\
Average exon length&$212\pm 5$(n=4885)&$240\pm 2$(n=62851)&$9\times 10^{-5}$\\
&672&1365& \\
5' UTR length&$135\pm   8$        &$ 173\pm  3$         &$4\times 10^{-  7}$\\
& 79& 106& \\
    3' UTR length     &$599 \pm  30$        &$846 \pm 13$         &$3\times 10^{-13 }$\\
&333& 552& \\
Coding sequence length&$1211\pm  44$        &$1770\pm 26$         &$3\times 10^{-26 }$\\
&928&1322& \\
Number of introns      &$8.2 \pm 0.3$        &$10.6\pm 0.2$         &$6\times 10^{-7  }$\\
&6  &   8& \\
Intron bps per coding bp\ \ \ \ \ \ \ \ & $20  \pm   2$        &$31.8\pm 0.8$         &$2\times 10^{-11 }$\\
&9.9&15.6& 
\end{tabular}

\end {table}

Table 1 compares the lengths of various parts of the HK
genes and the background genes. The alignment data was
taken from the UCSC genome browser (http://genome.ucsc.edu) 
\cite{16}. We excluded 322 genes that do not have a
unique alignment, as well as 1242 genes that were not
expressed in any tissue (to avoid potential problems
because of defective probes). This left 532 HK genes and
5404 non-HK genes. The histograms in Fig. 2-4 compare
HK genes with the other genes by total intron length, 5'
UTR length and coding sequence length. Remarkably,
there was a statistically significant difference between HK
and non-HK genes in all aspects of gene structure. Average
intron length is shorter for the HK genes than for the
background genes (2573 bp versus 5025 bp, respectively);
total gene length is shorter (21,050 bp versus 53,418 bp);
average exon length is shorter (212 bp versus 240 bp);
average lengths of both 3' and 5' untranslated regions
(UTRs) are shorter (5': 135 bp versus 173 bp; 3': 599 bp
versus 846 bp); and, most notably, the translated proteins
are shorter as well (403 amino acids versus 590 amino
acids). Accordingly, the number of introns bp per unit
of coding sequence length is lower for the HK genes
(20 versus 32). We studied the structure of each gene as a
function of the number of tissues it is expressed in and
verified that the results are not due to bias of the non-HK
genes by tissue-specific genes (data not shown).

\begin{figure}[t]
\includegraphics[width=2.5in]{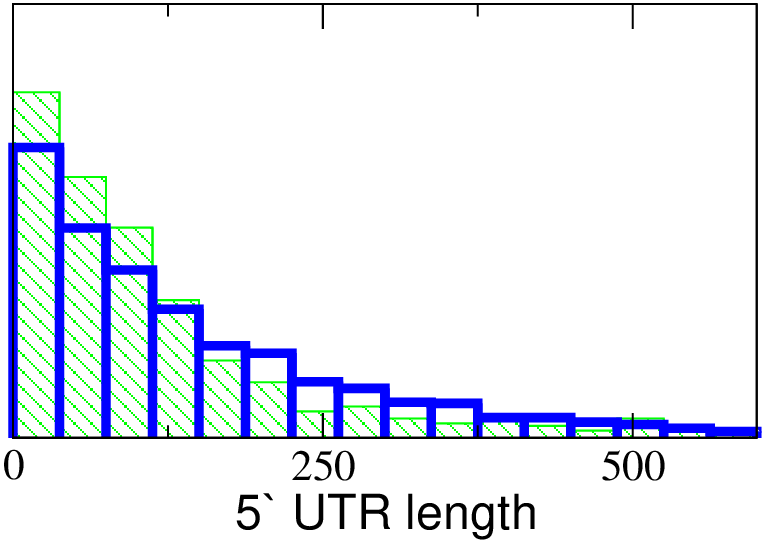}
\caption{A Histogram of the length of the 5' untranslated regions 
(UTR). Green bars, HK genes; blue bars, non-HK genes.}
\end{figure}

The pronounced statistical characteristics of the HK
gene set further supports their assignment as a unique set.
Our findings confirm and extend previous research,
showing that the introns of highly expressed genes are
shorter \cite{5}. As mentioned above, the HK genes expression
levels are high, and the fact that they have to be expressed
in all cells at all times makes them even more costly to
transcribe. Previously \cite{8}, the high abundance of a certain
gene in EST libraries was an indication the gene was
highly expressed in the human body. It was pointed out \cite{8},
however, that this method is prone to bias due to the
inclusion of normalized and tumor libraries and overrepresentation
of certain tissues. Our approach overcomes
this difficulty and confirms the previous result. Moreover,
we find here that UTRs and even the encoded proteins are
shorter for the HK genes. The magnitude of the difference
is greater for the introns than for the exons and proteins
(Table 1), which makes sense because the coding sequences
and the UTRs are less susceptible to change.

It should be mentioned that intronless genes were
included in our analysis after verifying that their inclusion
or exclusion had no effect on the results. It also must be
noted that the UTRs are not always fully sequenced, and
thus their actual lengths might be longer. This bias was
found to have no effect on the length of the coding
sequences, and in any case the effect would be the same
for both HK and non-HK genes.

It has been noted that codon usage bias in nonmammalian
organisms is correlated with the expression
level and with the gene length \cite{17,18,19}. These results led to
the conjecture of selective pressure on highly expressed
genes resulting in shorter proteins \cite{19}. However, no
evidence for this selection was found \cite{18}, possibly because
of a lack of high quality databases for these organisms.
Recent works have suggested that there is no selection for
codon usage bias in humans \cite{20}, and thus our results
demonstrate that the expression-length correlation is
not related to the expression-codon bias correlation.

\begin{figure}
\includegraphics[width=2.5in]{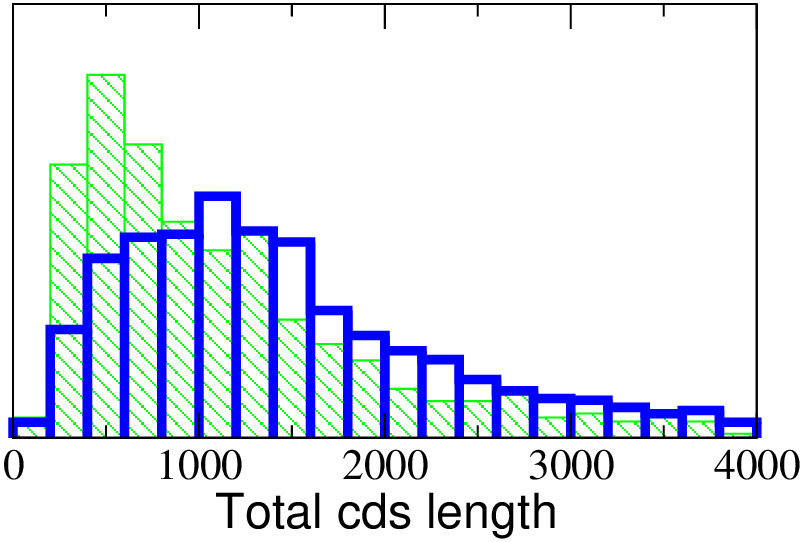}
\caption{A Histogram of the length of the coding region. Green bars, HK genes; 
blue bars, non-HK genes.}
\end{figure}

It could be argued that selection towards shorter genes
should have eliminated the introns in highly expressed
genes. However, it is known that introns do have
important roles, such as splicing regulation. Therefore,
there is a balance between the advantageous contribution
of the introns and the selective pressure for shortening.

Finally, when we compared our results with two (largely
overlapping) published sets of HK genes, we found that
roughly half of the genes in the intersection of those sets
were present in our set. We used the genomic structure to
test the remaining genes, and found a statistically
significant difference between them and our HK gene
set. The differences between our results and those of
earlier studies \cite{3,4} could be due to the fact that the
database we used was based on more advanced chip
technology and included many more different tissues,
giving it more discriminative power to identify HK genes.

In conclusion, we have identified a set of HK genes. The
set is publicly available at 
http://www.compugen.co.il/supp\_info/Housekeeping\_genes.html
and can be used for
calibration of microarrays, toxicity evaluation and quantitative
PCR experiments. Furthermore, we show that
HK genes have shorter introns, UTRs and coding
sequences, attesting to the strong selection for compactness
in these genes.

\begin{acknowledgments}
We thank Andrew Su for helpful discussion and for providing us with the
RefSeq mapping. Gady Cojocaru and Rotem Sorek are acknowledged for
comments on the manuscript and insightful discussion.
\end{acknowledgments}

\end{document}